\documentclass[aps,prl,superscriptaddress,twocolumn,showpacs,floatfix]{revtex4}

\usepackage{amsmath}
\usepackage{amssymb}
\usepackage{bm}
\usepackage{graphicx}

\begin{document}

%\draft

\title{Probing quantum-mechanical level repulsion in disordered
systems by means of time-resolved selectively-excited resonance
fluorescence}

\author{A.\ V.\ Malyshev}
\thanks{On leave from Ioffe Physiko-Technical Institute, 26
Politechnicheskaya str., 194021 St.-Petersburg, Russia}
\affiliation{GISC, Departamento de F\'{\i}sica de Materiales,
Universidad Complutense, E-28040 Madrid, Spain}

\author{V.\ A.\ Malyshev}
\thanks{On leave from V. A. Fock Institute of Physics, St.-Petersburg
University, 198904 St.-Petersburg, Russia} \affiliation{Institute
for Theoretical Physics and  Materials Science Centre, University of
Groningen, Nijenborgh 4, 9747 AG Groningen, The Netherlands}

\author{J.\ Knoester}
\affiliation{Institute for Theoretical Physics and  Materials
Science Centre, University of Groningen, Nijenborgh 4, 9747 AG
Groningen, The Netherlands}

\date{\today}

\begin{abstract}

We argue that the time-resolved spectrum  of selectively-excited
resonance fluorescence at low temperature provides a tool for
probing the quantum-mechanical level repulsion in the Lifshits tail
of the electronic density of states in a wide variety of disordered
materials. The technique, based on detecting the fast growth of a
fluorescence peak that is red-shifted relative to the excitation
frequency, is demonstrated explicitly by simulations on linear
Frenkel exciton chains.

\end{abstract}


\maketitle

Anderson localization is a key concept in understanding the
transport properties and optical dynamics of quasiparticles in
disordered systems in general~\cite{Abrahams79,Kramer93,Mirlin00}
and in a large variety of low-dimensional materials in particular.
Examples of current interest are conjugated
polymers~\cite{Hadzii99}, molecular
J-aggregates~\cite{Kobayashi96,Bednarz03}, semiconductor
quantum-wells~\cite{Alessi00,Klochikhin04} and quantum
wires~\cite{Feltrin04a}, as well as biological antenna
complexes~\cite{vanAmerongen00,Cheng06}. Localization results in the
appearance of a Lifshits tail in the density of states (DOS) below
the bare quasiparticle band~\cite{Lifshits88,Chernyak99}. At low
temperature, the states in the tail determine the system's transport
and optical properties. Of particular importance for the dynamics
and the physical properties is the spatial overlap between these
states. Two situations can be distinguished: states that do not
overlap can be infinitesimally close in energy, while states that do
overlap undergo quantum-mechanical level repulsion. Interestingly,
this repulsion does not manifest itself in the overall level
statistics, which, due to the localization, is of Poisson nature.
The local Wigner-Dyson statistics, caused by the repulsion, turns
out to be hidden under the global distribution of
energies~\cite{Malyshev01a}. Still, due to the characteristic energy
scale associated with the level repulsion, this phenomenon does
affect global properties, such as transport~\cite{Malyshev03}.
Moreover, observing changes in the level statistics allows one to
detect the localization-delocalization transition or mobility
edges~\cite{Mirlin00}.

Thus, it is of general interest to have experimental probes for the
level statistics in disordered systems. Recently, it has been shown
that near-field spectroscopy~\cite{Intonti01,Feltrin04b} and
time-resolved resonant Rayleigh scattering~\cite{Haacke01} may be
used to uncover the statistics of localized Wannier excitons in
disordered quantum wells~\cite{Intonti01} and
wires~\cite{Feltrin04b}. In this Letter, we argue that
low-temperature time-resolved selectively excited fluorescence from
the Lifshits tail provides an alternative tool for probing the level
repulsion. This method is based on the fact that downward relaxation
between spatially overlapping states dominates the early-time rise
of a fluorescence peak that is red-shifted relative to the
excitation frequency. The red shift is related to the level
repulsion. We will demonstrate this by simulations on a
one-dimensional (1D) Frenkel exciton model with on-site disorder.
The key ingredients of the model -- level repulsion and scattering
rates proportional to the phonon spectral density times the overlap
integral of the site probabilities of the two exciton states
involved~\cite{Bednarz03} -- are also characteristic for Wannier
exciton systems~\cite{Alessi00,Klochikhin04,Feltrin04a,Shimizu98}.
Hence, the method has a wide range of applicability.

Our model consists of an open chain of $N$ optically active
two-level units (monomers) with parallel transition dipoles,
coupled to each other by dipole-dipole interactions. The chain's
optical excitations are Frenkel excitons described by a
Hamiltonian matrix whose diagonal elements are the monomer
excitation energies, $H_{nn} = \varepsilon_n$, which are random
uncorrelated Gaussian variables with zero mean and standard
deviation $\Delta$ (i.e., $\langle \varepsilon_n \rangle = 0$ and
$\langle \varepsilon_n \varepsilon_m \rangle = \Delta^2
\delta_{nm}$, with $\langle \ldots \rangle$ denoting the average
over the disorder realizations). The off-diagonal matrix elements
are non-random dipole-dipole interactions: $H_{nm} = -
J/|n-m|^{3}$ \, $(J_{nn} \equiv 0)$, with $J> 0$ representing the
magnitude of the nearest-neighbor coupling. The exciton
eigenenergies $E_\nu$ and eigenfunctions $\varphi_{\nu n}$ follow
from diagonalizing $H_{nm}$.

\begin{figure}[ht]
\centerline{\includegraphics[width=4cm,angle=-90,clip]{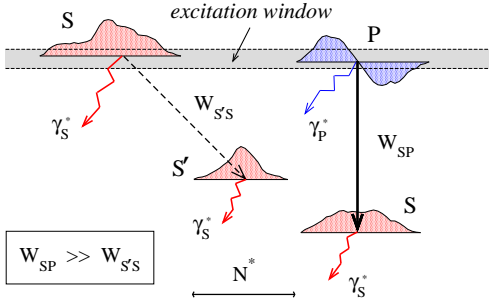}}
\caption{Cartoon of the local DOS tail states, illustrating the
possible channels of the exciton relaxation at zero temperature
after excitation within a narrow window. The solid (dashed) arrow
shows the intra (inter) segment relaxation with the rate
$W_\mathrm{SP}$ ($W_\mathrm{S^{\prime}S}$), while the wavy arrows
denote the radiative decay with the typical rate
$\gamma^*_{\mathrm{S}} = \gamma_0 N^*$ and $\gamma^*_{\mathrm{P}}
\approx 0.1\gamma_0 N^*$ ($N^*$ is the typical segment length).}
\label{fig0}
\end{figure}

The states in the Lifshits tail, which dominate the optical
absorption of the disordered chain, are localized on segments of
typical size $N^*$. Some of these states can be joined in {\em local
exciton manifolds} of a few states that are similar to the
eigenstates of an ideal chain of size equal to the segment
length~\cite{Malyshev01a,footnote1}. The states in a local manifold
undergo level repulsion, while between different segments energy
separations may be arbitrarily small. Because the variation of
exciton energies over different segments is larger than the level
repulsion within segments~\cite{Malyshev01a}, the latter remains
hidden in techniques that probe global averages like, e.g., the
absorption spectrum.

The typical situation is illustrated in Fig.~\ref{fig0}. The two
states to the right form a local manifold, with the lower state
having no node and collecting most of the oscillator strength of the
monomers within the segment. We will refer to this as an S type
state. The higher state has one node (P type) and has a much smaller
(but nonzero) oscillator strength; thus, it can be photo-excited but
its radiative decay is slow compared to that of the S state. The
probability overlap between the two states, measured by the integral
$I_{\mu\nu} = \sum_{n=1}^N \varphi_{\mu n }^2 \varphi_{\nu n}^2$, is
on the order of $1/N_{\mathrm {seg}}$, with $N_{\mathrm {seg}} \sim
N^*$ the segment (localization) size. The other two states in
Fig.~\ref{fig0} are the lowest (S type) states of other localization
segments. For linear systems, the overlap between states of adjacent
segments typically is two orders of magnitude smaller than the
overlap within a manifold~\cite{Malyshev01a}; for distant segments
the overlap is even much smaller. These overlap properties play a
crucial role in the intraband exciton relaxation.

To describe this relaxation, we use the Pauli master equation for
the populations $P_\nu$ of the exciton states:
\begin{equation}
    {\dot P}_\nu = R_\nu F_\nu -\gamma_\nu P_\nu + \sum_{\mu = 1}^N
    (W_{\nu\mu}P_{\mu} - W_{\mu\nu}P_\nu) \ ,
\label{Pnu}
\end{equation}
where $R_\nu$ is a source term, specified below, $F_\nu =
(\sum_{n=1}^N \varphi_{\nu n})^2$ is the dimensionless oscillator
strength of the $\nu$th exciton state, $\gamma_\nu = \gamma_0 F_\nu
$ is its spontaneous emission rate ($\gamma_0$ being the spontaneous
emission constant of a monomer), and $W_{\mu\nu}$ is the scattering
rate from the exciton state $|\nu\rangle$ to the state
$|\mu\rangle$, which results from weak exciton-vibration coupling.
The latter may be obtained through Fermi's golden
rule~\cite{Bednarz03}: $W_{\mu\nu} = I_{\mu\nu}\;
S(|\omega_{\mu\nu}|) \; G(\omega_{\mu\nu})$, where $I_{\mu\nu}$ is
the overlap integral defined above and $S(|\omega_{\mu\nu}|)$ is the
phonon spectral density at the energy $\omega_{\mu\nu} =
E_\mu-E_\nu$. Furthermore, $G(\omega)=n(\omega)$ if $\omega>0$ and
$G(\omega)=1+n(-\omega)$ if $\omega<0$, with
$n(\omega)=[\exp(\omega/T)-1]^{-1}$, the mean thermal occupation
number of a phonon mode of energy $\omega$ ($\hbar=k_B=1$).

We consider an experiment in which a short narrow-band laser pulse
is used to selectively excite states in a small part of the
absorption band and set $R_{\nu}(t) = R_0 \delta(t) \delta(E_l -
E_{\nu})$ with $E_l$ being the laser frequency. We are interested
in the time-resolved fluorescence spectrum at low temperature
($T=0$). The spectrum is given by
\begin{equation}
    I(E,t) = \frac{1}{N}\Big \langle\sum_{\nu=1}^N \gamma_\nu\, P_\nu(t)
    \> \delta(E - E_\nu) \Big\rangle \ .
\label{I}
\end{equation}
At $T = 0$, only states with $E_{\nu} \le E_l$ are relevant. Quite
generally, the spectrum consists of a narrow resonant peak,
resulting from fluorescence from initially excited states and a
red-shifted broader feature, arising from states that are populated
by downward relaxation from the initially excited ones.

If the laser frequency lies in the main part of the absorption band,
not too far in the blue wing, the initially excited states are
either of the S or of P type. Excitation of higher excited states,
which extend over various segments, may be neglected. Both S and P
states contribute to the narrow resonant fluorescence, with most of
the intensity coming from the S type states, because they carry more
oscillator strength. On the other hand, the P type states have a
much higher probability to relax to their S partner before they
decay radiatively. The S type states can only relax to states of
other segments, which is a slow process as compared to the
intra-manifold relaxation ($W_{\mathrm{S}{\mathrm P}} \gg
W_{\mathrm{S}^{\prime}{\mathrm S}}$). Therefore, at times $t <
1/W_{\mathrm{S}{\mathrm P}}$, any red-shifted fluorescence comes
from S type states that have been excited via intra-segment
relaxation from P type ones. Thus, the shape and position of the
red-shifted fluorescence band reflects the level repulsion, i.e.,
its intensity tends to zero when approaching the laser frequency. At
longer times, this repulsive gap may be partially filled due to
inter-segment hops.

We now analyze in more detail the early-time shape of the
red-shifted band $I_\mathrm{red}(E,t)$. For $t \ll
1/\gamma^*_{\mathrm{P}} + 1/W_{\mathrm{S}{\mathrm P}}$, one
iteration of the master equation gives
\begin{eqnarray}
    I_{\mathrm {red}}(E,t) & = & \frac{R_0 t}{N}
    \Big \langle\sum_{\mu,\nu}
    \gamma_\nu\, W_{\nu\mu}F_\mu \delta(E_\mu - E_l)
    \> \delta(E - E_\nu) \Big\rangle
    \nonumber\\
    & = & \frac{R_0 t}{N}S(E_l-E)\Big \langle
    \sum_{\mu,\nu}\gamma_\nu\, I_{\nu\mu}F_\mu\,
    \nonumber\\
    & \times & \delta(E_\mu - E_l)\> \delta(E - E_\nu) \Big\rangle \
    ,
    \label{Itsmall2}
\end{eqnarray}
where the sum is restricted to states which occur in SP doublets, as
described above. Thus, $\mu$ ($\nu$) in the sum label P (S) type
states. Because $I_{\mathrm {SP}} \sim 1/N_{\mathrm{seg}}$, while
the lower state in each segment has a superradiant decay rate
$\gamma_{\mathrm {S}} \approx \gamma_0 N_{\mathrm{seg}}$, we have
$\gamma_\nu\, I_{\nu\mu} = a \gamma_0$, with $a$ being a constant of
the order unity. If we now replace $F_\mu$ by the average oscillator
strength $F_{\mathrm P}(E_l)$ for P type states at the energy $E_l$,
we arrive at
\begin{equation}
    I_{\mathrm {red}}(E,t)  =
    \frac{a\gamma_0 R_0 t}{N}\> F_{\mathrm P}(E_l) \>S(E_l - E)\>
    P_{\mathrm{SP}}(E_l,E) \ .
\label{Itsmallfin}
\end{equation}
Here, $P_{\mathrm{SP}}(E_l,E)= \big\langle \sum_{\mu,\nu}
\delta(E_\mu - E_l)\> \delta(E - E_\nu) \big\rangle$ is the
probability distribution for the energy of an S state under the
condition that the same localization segment contains a P state at
the energy $E_l$. This result shows that the early-time lineshape is
the product of the phonon spectral density and the conditional
spacing distribution, thus containing detailed information about the
repulsive level statistics.

To confirm the above analytical result, we have performed numerical
simulations of the spectrum Eq.~(\ref{I}) for 1D Frenkel chains of
length $N=1000$ at $T=0$. In all calculations we considered system
parameters that are appropriate for linear aggregates of the dye
pseudoisocyanine (PIC) with counter ion F$^{-}$: $\gamma_0 = 2.7
\times 10^{8}\, {\mathrm{s}}^{-1} = 1.5\times 10^{-5}J$, $J = 600$
cm$^{-1}$, and $\Delta=0.25\, J$, giving a typical localization size
$N^* = 28$. For the spectral density we used the Debye model:
$S(\omega) = W_0 (\omega/J)^3$, with $W_0 = 15.9 J$. This set of
parameters has been used successfully to analyze and fit experiments
on aggregates of PIC-F~\cite{Bednarz03}. As at $T=0$ only states
below the excitation energy are relevant, we used the Lanczos method
to calculate the subset of only these eigenstates. The fluorescence
spectra were obtained after averaging over $10^5$ realizations of
the on-site energies.

\begin{figure}[ht]
\centerline{\includegraphics[width=6cm,clip]{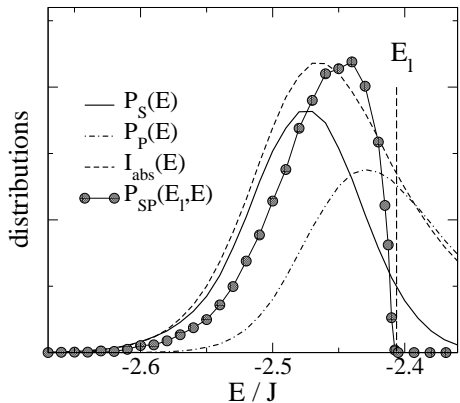}}
\caption{Calculated energy distributions of S type states (solid
curve), P type states (dash-dotted curve), absorption spectrum
(dashed curve), and the conditional probability
$P_{\mathrm{SP}}(E_l,E)$ (filled circles). } \label{fig2}
\end{figure}

First, we analyzed numerically various statistical properties of the
exciton levels. The results are summarized in Fig.~\ref{fig2}. Shown
are the distribution of energies of the S and P type states,
$P_{\mathrm S}(E)$ and $P_{\mathrm P}(E)$, respectively, the
absorption spectrum, $I_{\mathrm {abs}}(E)$, and the conditional
probability $P_{\mathrm{SP}}(E_l,E)$, with $E_l=-2.41J$. The
selection of doublets of S and P type states was done according to
the method described in
Ref.~\onlinecite{Malyshev01a}~\cite{footnote2}. Clearly, the
low-energy half of the absorption peak is dominated by S type
states, while the high-energy wing contains also important
contributions from P type states. Both distributions overlap, which
is the reason that the local energy structure is not seen in the
absorption peak. The conditional probability clearly demonstrates
the level repulsion, as it drops to zero at $E=E_l$. In our
simulations the laser energy $E_l$ was chosen such that it pumps
primarily P type states.

\begin{figure}[ht]
\centerline{\includegraphics[width=6cm,clip]
{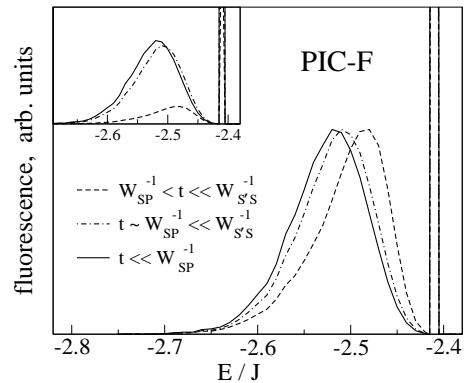}}
\caption{ Zero-temperature time-resolved fluorescence spectra at
different instants. The intensities of the red-shifted bands are
normalized to the same peak value. The inset shows the unnormalized
spectra.} \label{FLN}
\end{figure}

Figure~\ref{FLN} presents our numerical results for the
time-resolved fluorescence spectrum after selective excitation
within a window [$-2.415J, -2.405J$]. Indeed, the spectrum reveals a
sharp peak at the excitation energy and a clearly separated
red-shifted (R) band, indicative of level repulsion. The kinetics of
this band can be understood from estimates of the radiative decay
and relaxation rates. Typically, $\gamma_{\mathrm{S}}^* = \gamma_0
N^* \approx 10 \gamma_{\mathrm{P}}^*  \approx 4 \times 10^{-4} J$,
while $W_{\mathrm{SP}} \approx 10^2 W_{\mathrm{S^{\prime}S}} \approx
W_0 (\Delta^*/J)^3 I_{\mathrm{SP}} \approx 6 \times 10^{-5} J$. Here
we used the typical energy separation $\Delta^* = \Delta/\sqrt{N^*}$
and $I_{\mathrm{SP}} \approx 1/N^*$ ($N^* = 28$). Thus,
$W_{\mathrm{S^{\prime}S}} \ll \gamma_{\mathrm{P}}^* <
W_{\mathrm{SP}} \ll \gamma_{\mathrm{S}}^*$, implying that
inter-segment hops are irrelevant on the radiative time scale. As a
consequence, the observed blue-shift of the R band upon increasing
time, cannot be the result of filling the repulsive gap by such
hops. Instead, this kinetics results from the fact that the
oscillator strength per state peaks at lower energies than the
absorption maximum~\cite{Bednarz03}. Thus, the higher-energy states
in the R band decay slower, causing the blue-shift.

\begin{figure}[ht]
\centerline{\includegraphics[width=6cm,clip]
{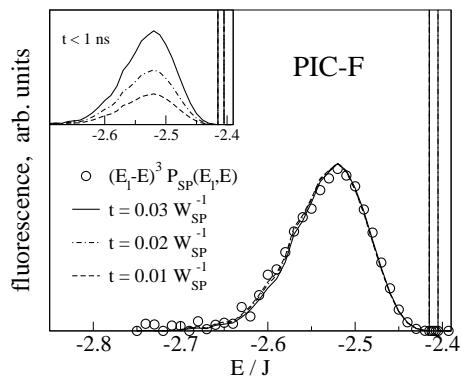}}
\caption{ Simulated fluorescence spectra at three early instances of
time, normalized by dividing by $t$ and compared to the function
$(E_l-E)^3 P_{\mathrm{SP}}(E_l,E)$ (open circles). The inset shows
the unnormalized  spectra.} \label{E21distribution}
\end{figure}

Finally, Fig.~\ref{E21distribution} displays the numerically
obtained early-time fluorescence spectra, for
$t=0.01W_{\mathrm{SP}}^{-1}$, $0.02W_{\mathrm{SP}}^{-1}$, and
$0.03W_{\mathrm{SP}}^{-1}$. The spectra are normalized by dividing
by the time $t$ at which they were taken. Clearly, the resulting
curves cannot be distinguished. Moreover, they agree almost
perfectly with the function $(E_l-E)^3 P_{\mathrm{SP}}(E_l,E)$,
plotted as open circles. This confirms the analytical
expression~(\ref{Itsmallfin}). Without showing explicit results,
we note that the same quality of agreement has been found for
other model parameters and spectral densities.

In summary, we have shown that time-resolved selective excitation of
fluorescence in the Lifshits tail of low-temperature disordered
systems, yields a probe for the quantum level repulsion and local
level statistics. Shortly after the laser pulse, the shape of the
red-shifted fluorescence band arising in this technique is the
product of the level spacing distribution within localization
segments and the phonon spectral density. To extract the spacing
distribution, the spectral density must be measured. The widely
adapted scheme for this also employs low-temperature selectively
excited fluorescence, but under steady-state conditions. It is
assumed that the red-shifted band directly reflects the phonon
spectral density~\cite{Renger02}. However, for multichromophore
systems with a fast intra-band relaxation, this approach is not
appropriate, because a relaxation-induced red-shifted feature, which
also arises under steady-state conditions~\cite{Bednarz05}, is
superimposed with the phonon side-band. The two contributions may be
separated by measuring the time-dependent spectrum. Thus, time
resolution is an essential aspect of the scheme proposed. Of course,
the relation between pulse width (selective excitation) and duration
(time-resolution) limits application of our approach to systems for
which the absorption band is wide compared to the intraband
relaxation rate. For the example considered here, this criterion is
easily met.

To conclude we stress that our method has a wider applicability than
the 1D Frenkel exciton system we modeled. The reason is that the key
ingredients -- the nature of the states in the Lifshits tail and the
importance of spatial overlap for their phonon-induced relaxation --
are shared by a large variety of
systems~\cite{Alessi00,Klochikhin04,Feltrin04a,Bednarz03,Shimizu98}.

This work was supported by the programs Ram\'on y Cajal (Ministerio
de Ciencia y Tecnolog{\'\i}a de Espa{\~n}a) and NanoNed (Dutch
Ministry of Economic Affairs).

%\clearpage
%\newpage


\begin{thebibliography}{99}
\bibitem{footnote1} The segments on which such local manifolds
are found usually do not correspond to the optimal disorder
fluctuations. The latter describe states that occur deeper in the
Lifshits tail (see, e.g., Ref.~\onlinecite{Chernyak99}), localized
as singlets, without higher-energy partners on the same segment.
\bibitem{footnote2} To select the SP doublets,
    we used $\mathrm{C}_1 = 0.9$ and $\mathrm{C}_2 = 0.7$ in the
    selection rules elaborated in Ref.~\onlinecite{Malyshev01a}.
\end{thebibliography}
\end{document}